\documentstyle[preprint,aps,epsf,floats]{revtex}

\tighten

\def\OMIT#1{{}}

\def\si{{}^1\kern-.14em S_0}
\def\siii{{}^3\kern-.14em S_1}
\def\piii{{}^3\kern-.14em P_1}
\def\diii{{}^3\kern-.14em D_1}

\begin{document}

\preprint{\vbox{
\hbox{NT@UW-02-002}
}}

\title{Nucleon-Nucleon Interactions on the Lattice}
\author{{\bf Silas R.~Beane}  and {\bf Martin J.~Savage}}
\address{Department of Physics, University of Washington, \\
Seattle, WA 98195. }

\maketitle

\begin{abstract} 
  We consider the nucleon-nucleon potential in quenched and partially-quenched
  QCD. The leading one-meson exchange contribution to the
  potential is found to fall off exponentially at long-distances, in contrast
  with the Yukawa-type behaviour found in QCD. 
This unphysical component of the two-nucleon potential has important 
implications for the extraction of
  nuclear properties from lattice simulations. 
\end{abstract}

\bigskip
\vskip 8.0cm
\leftline{February 2002}

\vfill\eject


It is hoped that lattice methods will ultimately lead to an understanding of
nuclear physics directly from QCD.  To date there exist two distinct approaches
to investigating the nucleon-nucleon (NN) system using lattice QCD. The first
approach makes use of L\"uscher's finite-volume algorithm, which expresses the
energy of a two-particle state as a perturbative expansion in the scattering
length divided by the size of the box~\cite{Luscher}. Practical computations
require a sufficiently small box, which in turn should be larger than the
scattering length. Hence L\"uscher's method is ideal for systems with
``natural'' scattering lengths -- with size of order a characteristic physical
length scale -- and has been used to study $\pi\pi$ 
scattering~\cite{Gupta:1993rn}.
Of course in the NN system, the scattering lengths are much
larger than the characteristic physical length scale given by
$\Lambda_{\scriptstyle QCD}^{-1}$. Nevertheless one may expect that at the
unphysical values of the pion mass currently used in lattice simulations, the
scattering lengths relax to natural values, thus allowing their determination
from the lattice. Attempts have been made to compute NN scattering parameters
in lattice QCD; Ref.~\cite{fuku} computes the $\si$ and $\siii$ scattering
lengths in quenched QCD (QQCD) using L\"uscher's method. A second,
less-ambitious approach to the NN system on the lattice is to study the
simplified problem of two interacting heavy-light
particles~\cite{Richards:1990xf,Mihaly:1996ue,Stewart:1998hk,Michael:1999nq,Richards:2000ix}.
Here the hope is to extract information about the long-distance tail of the
potential, which in nature is provided by one-pion exchange (OPE).

At present computational efforts are limited by the use of quark masses
that are significantly larger than those of nature, typically 
$m_\pi^{\rm   latt.}\sim 500~{\rm MeV}$. In order to make a connection between 
present (and near-future) lattice calculations and nature, an extrapolation to 
the physical values of the quarks masses is required.  
The technology which allows a
systematic extrapolation has been developed for 
QQCD~\cite{Sharpe90,S92,BG92,LS96,S01a} and partially quenched QCD
(PQQCD)~\cite{Pqqcd,SS01,CS01a,BS02a}.
In this paper we will consider the potential between two
nucleons at leading order (LO) in quenched chiral
perturbation theory (Q$\chi$PT) and in partially-quenched chiral
perturbation theory (PQ$\chi$PT),
the low-energy effective field theories that describe QQCD and 
PQQCD, respectively.


We will be considering the extension from two-flavor QCD,
where the three $\pi$'s are pseudo-Goldstone bosons associated with the 
breaking of chiral symmetry and the $\eta$ is the $SU(2)$ singlet meson
whose mass arises mainly from the $U(1)_A$ axial anomaly,
to QQCD and PQQCD.
Essential to our discussion will be the modifications to the
$\eta$ propagator which result from quenching and partial quenching. 
In QQCD the quarks, $Q=(u,d,{\tilde u},{\tilde d})^T$, live in the
fundamental representation of $SU(2|2)$, with  
the bosonic quarks, ${\tilde u},{\tilde d}$, 
accounting for the graded structure of the symmetry group.
The low-energy effective field theory (EFT)
is Q$\chi$PT, whose degrees of freedom are constrained by the chiral group 
$SU(2|2)_L\otimes SU(2|2)_R$.
In the isospin limit 
the $\eta$ propagator in Q$\chi$PT is given by~\cite{Sharpe90,S92,BG92}
\begin{eqnarray}
G_{\eta\eta} (q^2) & = & 
{i\over {(q^2- m_\pi^2 +i\epsilon )}}
\ +\ 
{{i(M_0^2-\alpha_\Phi q^2)}\over{(q^2- m_\pi^2 +i\epsilon )^2}}
\ \ .
\label{eq:etaqqcd}
\end{eqnarray}
The hairpin parameters $M_0$ and $\alpha_\Phi$ characterize the 
dynamics of the singlet field and do not vanish in the chiral limit. 
The double-pole, or hairpin, structure is an unphysical artifact
of quenching. Unfortunately QQCD is not related to QCD by any
smooth variation of parameters. This is not the case in PQQCD
where the quarks, (four fermionic and two bosonic) 
$Q=(u,d,j,l,{\tilde u},{\tilde d})^T$ live in the
fundamental representation of the graded symmetry group $SU(4|2)$ and the
low-energy EFT is PQ$\chi$PT, whose degrees of freedom are constrained by
chiral $SU(4|2)_L\otimes SU(4|2)_R$.
As the $j,l$ sea quarks become degenerate with the $u,d$ valence quarks,
PQQCD smoothly maps to QCD. If the sea quark masses are taken to infinity,
PQQCD formally goes to QQCD. In the isospin limit for both valence 
and sea quarks the $\eta$ propagator in PQ$\chi$PT is
\begin{eqnarray}
G_{\eta\eta} (q^2) & = & 
{{i (m_{\eta_j}^2-m_\pi^2)}\over{(q^2- m_\pi^2 +i\epsilon )^2}}
\ \ ,
\label{eq:etapqqcd}
\end{eqnarray}
where $m_{\eta_j}$ is the mass of the pseudo-Goldstone 
bosons composed of sea quarks.
Clearly this unphysical double pole vanishes in the QCD limit.
The hairpin structure resulting from (partial) quenching 
will have important
implications for lattice simulations of heavy-heavy systems,
such as the two-nucleon system or systems composed of two (or more) 
hadrons containing heavy quarks.


\begin{figure}[t]
\centerline{{\epsfxsize=4.0in \epsfbox{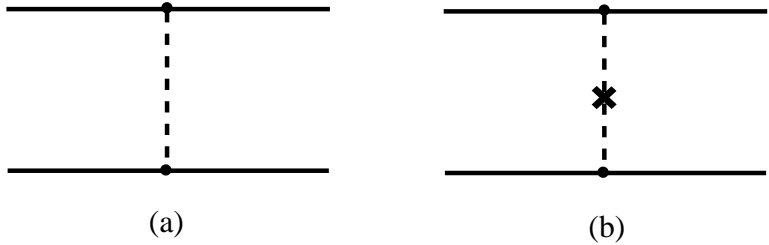}}} 
\vskip 0.15in
\noindent
\caption{\it 
Pion and $\eta$ exchange contributions to the
NN potential. 
The cross signifies a hairpin propagator 
from eq.~(\ref{eq:etaqqcd}) (QQCD) or  eq.~(\ref{eq:etapqqcd}) (PQQCD).}
\label{fig:NNpotential}
\vskip .2in
\end{figure}
\begin{figure}[t]
\centerline{{\epsfxsize=4.5in \epsfbox{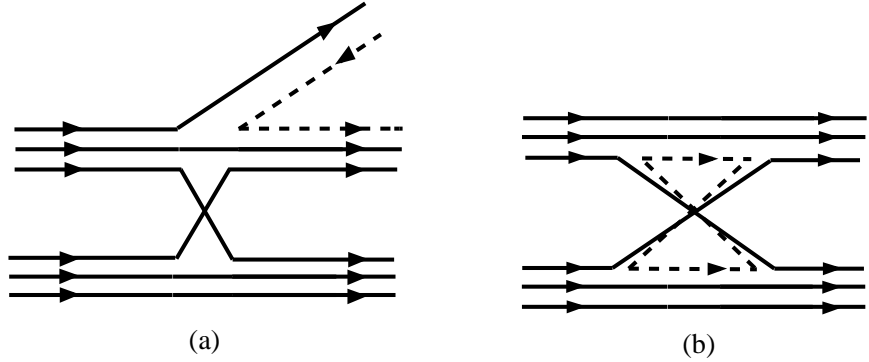}}} 
\vskip 0.15in
\noindent
\caption{\it The solid lines indicate valence quarks 
and the dashed lines
  indicate either a sea quark or a ghost quark. 
In the left panel (a) we
  exhibit a higher fock component of the NN potential 
and in the right panel (b)
  we exhibit a crossed-box loop correction to the 
NN potential. Both graphs are
  subleading in the EFT expansion.}
\label{fig:NNpotential2}
\vskip .2in
\end{figure}

The EFT power-counting for NN scattering has been developed in
Ref.~\cite{We90,KSWb,Beane:2001bc}. While details of the power-counting are
channel dependent, LO  for the low partial waves consists of OPE
and a four-nucleon contact interaction with no derivatives
or insertions of the quark mass matrix. This contact operator encodes
information about short-distance physics. 
The EFT power-counting is readily generalized to QQCD and PQQCD, and 
there is no direct\footnote{We define direct (partial) quenching to be 
(partial) quenching
beyond modifications to the singlet propagator.} (partial) 
quenching of the NN potential at LO in the EFT.
This is easily seen by assigning
a conserved charge to the various quark types.
We will consider PQQCD but the same argument applies to QQCD.
For instance, consider the charge assignments
$q_V=(1/3,0,0)$, $q_S=(0,1/3,0)$,
$q_G=(0,0,1/3)$
for valence-, sea- and ghost-quark number, respectively.
Here $q_V$ plays the role of ordinary baryon number. 
The NN initial state carries total quark-number charge $(2,0,0)$.
Baryon states with two valence quarks and either one ghost quark or
one sea quark carry charge $(2/3,0,1/3)$ or $(2/3,1/3,0)$, respectively, 
while meson states with one valence quark and
either one ghost or one sea anti-quark carry charge $(1/3,0,-1/3)$ or
$(1/3,-1/3,0)$, respectively. 
It follows
that the NN potential must have NN in the initial and in the final state.
Of course the potential is directly (partially) quenched, but only through loops,
which appear beyond LO in the EFT, as in
fig.~(\ref{fig:NNpotential2}b). Direct (partial) 
quenching modifies the tree-level potential through higher fock
components. For instance, one can
have a final state of one nucleon and one baryon composed of 
two valence quarks and one sea
or ghost quark and one meson composed of one valence quark and 
one sea or ghost
anti-quark. This contribution can arise either from meson-exchange graphs, 
as in 
fig.~(\ref{fig:NNpotential2}a), or from four-nucleon contact operators with one
insertion of the quark mass matrix or one insertion of a chiral covariant
derivative. In either case, the external meson legs at the level of the
potential must be reabsorbed by the final state nucleons. In contrast
with the ``potential'' pions exchanged between the baryons, these
``radiation'' pions generate contributions which are subleading in the EFT.
Hence at LO  there is no direct (partial) quenching
and the NN potential is given by the one meson exchange graphs of fig.~(\ref{fig:NNpotential}).
However, there are hairpin contributions to the potential at LO. 
For simplicity we work in the isospin limit for both the valence 
and the sea quarks. 
The diagrams in fig.~(\ref{fig:NNpotential}) give the LO 
meson-exchange contributions to the NN potential.
In coordinate space the NN potential is
\begin{eqnarray}
V^{(Q)} (r)\ =\ 
{1\over 8\pi f^2}\ 
{\bf\sigma}_1\cdot {\bf\nabla} {\bf\sigma}_2\cdot {\bf\nabla} 
\left( g_A^2\  {{\bf\tau}_1\cdot{\bf\tau}_2\over r}
\ +\ g_0^2\ {1-\alpha_\Phi\over r}
\ -\ g_0^2\ 
{M_0^2 - \alpha_\Phi m_\pi^2\over 2 m_\pi}
\right) \ e^{-m_\pi r} \ \ ,
\end{eqnarray}
in QQCD, where the axial coupling to the pions is $g_A$, while the 
coupling to the $\eta$ is $g_0$.
However, one should keep in mind that there is no relation between 
axial couplings in QQCD and those of QCD.
In PQQCD, we find that
\begin{eqnarray}
V^{(PQ)} (r)\ =\ 
{1\over 8\pi f^2}\ 
{\bf\sigma}_1\cdot {\bf\nabla} {\bf\sigma}_2\cdot {\bf\nabla} 
\left( g_A^2\ {{{\bf\tau}_1\cdot{\bf\tau}_2}\over r} -
g_0^2\ {{\left(m_{\eta_j}^2-m_\pi^2\right)}\over{2 m_\pi}}
\right) \ e^{-m_\pi r} 
\ \ ,
\end{eqnarray}
where the couplings $g_A$ and $g_0$ in this potential should be perturbatively
close to those of QCD.
Notice that effects due to partial-quenching (but not quenching)
disappear in the
short-distance limit, and are irrelevant to the renormalization of
the potential. This in turn implies that the four-nucleon contact operator,
that appears in PQQCD at the same order in the EFT as OPE, is that of
QCD~\cite{We90,KSWb,Beane:2001bc}. 
In contrast, the long-distance
behaviour of the potentials are dominated by the hairpin 
contributions,
\begin{eqnarray}
V (r)\ \rightarrow\ 
-{g_0^2\over 16\pi f^2}\ 
{\bf\sigma}_1\cdot \hat{\bf r} {\bf\sigma}_2\cdot \hat{\bf r}\ m_\pi 
\ e^{-m_\pi r} \  
\cases{ \left( M_0^2 - \alpha_\Phi m_\pi^2 \right)\ ,
& {\rm QQCD};\cr
        \left(m_{\eta_j}^2-m_\pi^2\right)\ ,& {\rm PQQCD}
\ \ \ ,\cr} 
\end{eqnarray}
which has both a central piece and a tensor piece.
Both quenching and partial-quenching
lead to an NN potential which falls off {\it exponentially} 
at long distances, as opposed to the physical Yukawa fall-off. 
In nature there is a fine-tuning between the long-distance Yukawa part of the
potential due to OPE, and the short-distance part which is
encoded in the contact operator which appears at the same order 
in the EFT.  
It is this fine-tuning which leads, 
for instance, to a shallowly bound deuteron.
It is clear that this fine-tuning does not exist for QQCD, and
quickly disappears in PQQCD
as one moves away from the QCD limit.


In conclusion, multi-nucleon quenched and partially-quenched simulations on the
lattice are complicated by the presence of a long-distance exponential tail,
that disturbs a fine-tuning which is a key feature of nuclear physics.  It has
been shown in Ref.~\cite{Beane:2001bc} that an understanding of the quark-mass
dependence of the deuteron binding energy requires the matrix element of an
operator that is not constrained by NN scattering data but which in principle
can be obtained from lattice QCD.  Any meaningful determination of this matrix
element from a partially-quenched simulation will have to account for the
unphysical, long-distance behavior highlighted in this work.

\bigskip\bigskip

\acknowledgements

This work is supported in
part by the U.S. Dept. of Energy under Grant No.  DE-FG03-97ER4014.


\begin{references}

\bibitem{Luscher}
M.~L\"uscher,
{\it Commun. Math. Phys.} {\bf 105}, 153 (1986).

\bibitem{Gupta:1993rn}
R.~Gupta, A.~Patel and S.~R.~Sharpe,
{\it Phys. Rev.} {\bf D48}, 388 (1993).

\bibitem{fuku}
M.~Fukugita, Y.~Kuramashi, M.~Okawa, H.~Mino and A.~Ukawa,
{\it Phys. Rev.} {\bf D52}, 3003 (1995).

\bibitem{Richards:1990xf}
D.~G.~Richards, D.~K.~Sinclair and D.~W.~Sivers,
{\it Phys. Rev.} {\bf D42}, 3191 (1990).

\bibitem{Mihaly:1996ue}
A.~Mihaly, H.~R.~Fiebig, H.~Markum and K.~Rabitsch,
{\it Phys. Rev.} {\bf D55}, 3077 (1997).

\bibitem{Stewart:1998hk}
C.~Stewart and R.~Koniuk,
{\it Phys. Rev.}  {\bf D57}, 5581 (1998).

\bibitem{Michael:1999nq}
C.~Michael and P.~Pennanen  [UKQCD Collaboration],
{\it Phys. Rev.} {\bf D60}, 054012 (1999).

\bibitem{Richards:2000ix}
D.~G.~Richards,
{\tt nucl-th/0011012}.

\bibitem{Sharpe90}
S. R. Sharpe,
{\it Nucl. Phys.}  {\bf B17} (Proc. Suppl.), 146 (1990).

\bibitem{S92}
S. R. Sharpe, 
{\it Phys. Rev.} {\bf D46}, 3146 (1992). 

\bibitem{BG92}
C. Bernard and M. F. L. Golterman,
{\it Phys. Rev.}  {\bf D46}, 853 (1992).

\bibitem{LS96}
J. N. Labrenz and S. R. Sharpe,
{\it Phys. Rev.}  {\bf D54}, 4595 (1996).

\bibitem{S01a}
M. J. Savage,
{\tt nucl-th/0107038}.


\bibitem{Pqqcd}
S.~R.~Sharpe and N.~Shoresh,
{\it Phys. Rev.} {\bf D62}, 094503 (2000);
{\it Nucl. Phys. Proc. Suppl.} {\bf 83}, 968 (2000);
M. F. L. Golterman and K.-C. Leung,
{\it Phys. Rev.} {\bf D57}, 5703 (1998).
S.~R.~Sharpe,
{\it Phys. Rev.} {\bf D56}, 7052 (1997);
C.~W.~Bernard and M.~F.~L.~Golterman,
{\it Phys. Rev.} {\bf D49}, 486 (1994).

\bibitem{SS01}
S.~R.~Sharpe and N.~Shoresh,
{\tt hep-lat/0011089};
{\it Phys. Rev.} {\bf D64}, 114510 (2001).

\bibitem{CS01a}
J. -W. Chen and M. J. Savage,
{\tt hep-lat/0111050}.

\bibitem{BS02a}
S.~R.~Beane and M.~J.~Savage,
{\it in preparation}.

\bibitem{We90}
S.~Weinberg,
{\it Phys. Lett.} {\bf B251}, 288 (1990);
{\it Nucl. Phys.} {\bf B363}, 3 (1991).

\bibitem{KSWb}
 D.~B.~Kaplan, M.~J.~Savage, and M.~B.~Wise, 
{\it Phys. Lett.} {\bf B424}, 390 (1998);
{\it Nucl. Phys.} {\bf B534}, 329 (1998).

\bibitem{Beane:2001bc}
S.~R.~Beane, P.~F.~Bedaque, M.~J.~Savage and U.~van Kolck,
{\tt nucl-th/0104030}.

\end{references}
\end{document}